\documentclass[11pt]{article}

\usepackage{latexsym}
\usepackage{amssymb}
\usepackage{euscript}
\usepackage{amsfonts}

\newcommand{\linimpl}{- \!\! \circ \,}

\newcommand{\isoarrow}{\stackrel{ \cong}{\longrightarrow}}

\newcommand{\CC}{\EuScript{C}}

\newcommand{\Cpo}{\ensuremath{\mathbf{Cpo}}}
\newcommand{\ident}{\ensuremath{\mathtt{id}}}

\newcommand{\Set}{\ensuremath{\mathbf{Set}}}

\newcommand{\YY}{\ensuremath{\mathbf{Y}}}

\newcommand{\fst}{\mathsf{fst}}
\newcommand{\snd}{\mathsf{snd}}

\newcommand{\restrict}{\upharpoonright}

\newcommand{\GG}{\mathcal{G}}
\newcommand{\labarrow}[1]{\stackrel{#1}{\longrightarrow}}
\newcommand{\Pow}{\mathcal{P}}
\newcommand{\MM}{\mathcal{M}}
\newcommand{\Res}{\mathcal{R}}
\newcommand{\ie}{\textit{i.e.}}
\newcommand{\cf}{\textit{cf.}}
\newcommand{\Tr}[3]{\mathtt{Tr}_{#1 , #2}^{#3}}
\newcommand{\Pfn}{\mathbf{Pfn}}
\newcommand{\PInj}{\mathbf{PInj}}
\newcommand{\Rel}{\mathbf{Rel}}
\newcommand{\inl}{\mathtt{inl}}
\newcommand{\inr}{\mathtt{inr}}
\renewcommand{\fst}{\mathtt{fst}}
\renewcommand{\snd}{\mathtt{snd}}
\newcommand{\sym}{\mathtt{sym}}

\newcommand{\GI}{\mathcal{GI}}

\begin{document}
\bibliographystyle{alpha}
\pagestyle{empty}
\title{Retracing some paths in Process Algebra}
	
\author{
	Samson Abramsky \\
	Laboratory for the Foundations of Computer Science \\
	University of Edinburgh	
}
\date{}
\maketitle

\section{Introduction}
The very existence of the \textsc{concur} conference bears witness to 
the fact that ``concurrency theory'' has developed into a subject unto 
itself, with substantially different emphases and techniques to those 
prominent elsewhere in the semantics of computation.

Whatever the past merits of this separate development, it seems 
timely to look for some convergence and unification. In addressing 
these issues, I have found it instructive to trace some of the 
received ideas in concurrency back to their origins in the early 1970's. 
In particular, I want to focus on a seminal paper by Robin Milner 
\cite{Mil75}\footnote{Similar ideas appeared independently in the work 
of Hans Beki\'{c} \cite{Bek71}.}, which 
led in a fairly direct line to his enormously influential work on 
\textsc{ccs} \cite{Mil80,Mil89}. I will take (to the extreme) the 
liberty of of applying hindsight, and show how some different paths 
could have been taken, which, it can be argued, lead to a more unified 
approach to the semantics of computation, and moreover one which may 
be better suited to modelling today's concurrent, object-oriented 
languages, and the type systems and logics required to support such 
languages.

\section{The semantic universe: transducers}
Milner's starting point was the classical automata-theoretic notion 
of \emph{transducers}, \ie\ structures
\[ (Q, X, Y, q_{0}, \delta ) \]
where $Q$ is a set of states, $q_{0} \in Q$ the initial state, $X$ 
the set of inputs, $Y$ the set of outputs, and
\[ \delta : Q \times X \rightharpoonup Y \times Q \]
is the transition function (here a partial function). If we supply a 
sequence of inputs $x_{0}, \ldots , x_{k}$ to such a transducer, we 
obtain the orbit
\[ q_{0} \labarrow{x_{0}} y_{0}, q_{1} \labarrow{x_{1}} y_{1}, q_{2} 
\labarrow{x_{2}} \cdots \labarrow{x_{k}} y_{k}, q_{k+1} \]
if $\delta (q_{i}, x_{i}) = y_{i}, q_{i+1}$, $0 \leq i \leq k$.
This generalizes to non-deterministic transducers with transition 
function
\[ \delta : Q \times X \longrightarrow \Pow (Y \times Q) \]
in an evident fashion.

The key idea in \cite{Mil75} is to give a denotational semantics for 
concurrent programs as \emph{processes}, which were taken to be 
extensional versions of transducers. There are two ingredients to this 
idea:
\begin{enumerate}
	\item  Instead of modelling programs by functions or relations, to 
	model them by entities with more complex behaviours, taking account 
	of the possible interactions between a program and its environment 
	during the course of a computation.
	\begin{quotation}
		``The meaning of a program should express its history of access to 
		resources which are not local to it.'' \cite{Mil75}
	\end{quotation}

	\item  Instead of modelling concurrent programs by automata, with 
	all the intensionality this entails, to look for a more extensional 
	description of the \emph{behaviours} of transducers.
\end{enumerate}
To obtain this extensional view of transducers, consider the 
recursive definition
\[ R = X \rightharpoonup Y \times R . \]
This defines a mathematical space of ``resumptions'' in which the 
states of transducers are ``unfolded'' into their observable 
behaviours. Milner solved equations such as this over a category of 
domains in \cite{Mil75}, but in fact it can be solved in a canonical 
fashion over $\Set$---in modern terminology, the functor 
\[ T_{X,Y} : \Set \longrightarrow \Set \]
\[ T_{X, Y} (S) = X \rightharpoonup Y \times S \]
has a final coalgebra $R \isoarrow T_{X, Y} (R)$. Indeed, Milner defined 
a notion $\sim$ of 
behavioural equivalence between transducers, and for any transducer 
$(Q, X, Y, q_{0}, \delta )$ a map $h_{\delta} : Q \longrightarrow R$ which is 
in fact the final coalgebra homomorphism from the coalgebra 
\[\hat{\delta} : Q \longrightarrow T_{X, Y} (Q) \]
to $R$ (where $\hat{\delta}$ is the exponential transpose of $\delta$), 
and proved that 
\[ (Q, X, Y, q_{0}, \delta ) \sim (Q', X, Y, q_{0}', 
\delta' ) \;\; \Longleftrightarrow \;\; h_{\delta}(q_{0}) = h_{\delta'}(q_{0}' ).
\]
From a modern perspective, we can also make light of a technical 
problem which figured prominently in \cite{Mil75}, namely how to 
model non-determinism. Historically, this called forth Plotkin's work 
on powerdomains \cite{Plo77a}, but for the specific application at 
hand, the equation
\[ R = X \longrightarrow \Pow (Y \times R) \]
has a final coalgebra in the category of classes in Peter Aczel's 
non-well-founded set theory \cite{Acz}, and if we are 
content to bound the cardinality of subsets by an inaccessible 
cardinable $\kappa$, then the equation
\[ R = X \longrightarrow \Pow^{< \kappa} (Y \times R) \]
has a final coalgebra in $\Set$ \cite{Bar}. Moreover, the equivalence 
induced by this model coincides with strong bisimulation \cite{Acz}.

However, this is not central to our concerns here. Rather, we want to 
focus on three important choices in the path followed by Milner from 
this starting point:
\begin{itemize}
	\item  Type-free \emph{vs.} typed

	\item  Extrinsic \emph{vs.} intrinsic interaction

	\item  Names \emph{vs.} information paths.
\end{itemize}
We want to examine the consequences of making different choices on 
these issues.
\subsection{Typed vs. type-free}
Rather than looking at a single type-free space of resumptions as 
above, and trying to invent some plausible operations on this space, 
we will focus instead on the \emph{category} of resumptions, and try 
to identify the structure naturally present in this category.

The category $\Res$ of resumptions (we will for simplicity confine 
ourselves to the deterministic resumptions) has as objects sets, and 
as morphisms
\[ \Res (X, Y) = X \rightharpoonup Y \times \Res (X, Y) \]
\ie\ the space of resumptions parameterized by the sets of ``inputs'' 
$X$ and ``outputs'' $Y$.
The composition of resumptions $f \in \Res (X, Y)$ and $g \in \Res (Y, 
Z)$ is defined (coinductively \cite{Acz}) by:
\[ f ; g(x) = \left\{ \begin{array}{ll}
              (z, f' ; g' ) &  f(x) = (y, f' ), \, g(y) = (z, g' ) \\
              \mbox{undefined} & \mbox{otherwise.}
              \end{array}
              \right. 
              \]
The identity resumption $\ident_{X} \in \Res (X,X)$ is defined by
\[ \ident_{X} (x) = (x, \ident_{X} ). \]
We can picture this composition as sequential (or ``series'') 
composition of transducers.

We can define a monoidal structure on $\Res$ by
\[ X \otimes Y = X + Y  \;\;\; \mbox{(disjoint union of sets)} \]
and if $f \in \Res (X, Y)$, $g \in \Res (X', Y')$, $f \otimes g \in 
\Res (X \otimes X' , Y \otimes Y' )$ is defined by:
\[ f \otimes g (\inl (x) ) = \left\{ \begin{array}{ll}
                              (\inl (y), f' \otimes g), & f(x) = (y, 
                              f') \\
                              \mbox{undefined} & \mbox{otherwise}
                              \end{array} \right.
\]
\[ f \otimes g (\inr (x') ) = \left\{ \begin{array}{ll}
                              (\inr (y'), f \otimes g'), & g(x') = (y', 
                              g') \\
                              \mbox{undefined} & \mbox{otherwise.}
                              \end{array} \right.
\]
This is (asynchronous) parallel composition of transducers: at each stage, we 
respond 
to an input on the $X$ ``wire'' according to $f$, with output appearing 
on the $Y$ wire, and to an input on the $X'$ wire according to $g$, 
with output appearing on the $Y'$ wire.

The remaining definitions to make this into a symmetric monoidal 
structure on $\Res$ are straightforward, and left to the reader.
Note that the associativity and symmetry isomorphisms, like the 
identities, have just one state; they are ``history-free''.

Finally, there is a feedback operator: for each $X$, $Y$, $U$ a 
function
\[ \Tr{X}{Y}{U} : \Res (X \otimes U, Y \otimes U) \longrightarrow 
\Res (X,Y) \]
defined by
\[ \Tr{X}{Y}{U}(f)(x) = \left\{ \begin{array}{ll}
                        (y, f'), & \exists k. \, f(x) = (u_{0}, 
                        f_{0}), \\
                        & f_{0}(u_{0}) = (u_{1}, f_{1}), \\
                        & \vdots \\
                        & f_{k}(u_{k}) = (y, f') \\
                        \mbox{undefined} & \mbox{otherwise.}
                        \end{array} \right.
\]
One should picture a token entering at the $X$ wire, circulating $k$ 
times around the feedback loop at the $U$ wire, and exiting at $Y$.

This feedback operator satisfies a number of algebraic properties (to 
simplify the statement of these properties, we elide associativity 
isomorphisms, \ie\ we pretend that $\Res$ is \emph{strict} monoidal):
\begin{description}
	\item[Naturality in $X$]  
	 \[ \Tr{X}{Y}{U}((g \otimes \ident_{U});f) = g ; \Tr{X'}{Y}{U}(f) \]
	 where $f : X' \otimes U \longrightarrow Y \otimes U$, $g : X 
	 \longrightarrow X'$.

	\item[Naturality in $Y$]  
	\[ \Tr{X}{Y}{U}(f; (g \otimes \ident_{U})) = \Tr{X}{Y'}{U}(f) ; g \]
	where $f : X \otimes U \longrightarrow Y' \otimes U$, $g : Y' 
	\longrightarrow Y$.

	\item[Naturality in $U$]  
	\[ \Tr{X}{Y}{U}(f;(\ident_{Y} \otimes g)) = 
	\Tr{X}{Y}{U'}((\ident_{X} \otimes g);f) \]
	where $f : X \otimes U \longrightarrow Y \otimes U'$, $g : U' 
	\longrightarrow U$.

	\item[Vanishing]  
	\[ \Tr{X}{Y}{I}(f) = f \]
	where $f : X \longrightarrow Y$, and
	\[ \Tr{X}{Y}{U \otimes V}(f) = \Tr{X}{Y}{U}(\Tr{X \otimes U}{Y 
	\otimes U}{V}(f)) \]
	where $f : X \otimes U \otimes V \longrightarrow Y \otimes U \otimes 
	V$.

	\item[Superposing]  
	\[ \Tr{X \otimes Z}{Y \otimes W}{U}((\ident_{X} \otimes 
	\sym_{Z,U}) ; (f \otimes g) ; (\ident_{Y} \otimes \sym_{U, W})) = 
	\Tr{X}{Y}{U}(f) \otimes g \]
	where $f : X \otimes U \longrightarrow Y \otimes U$, $g : Z 
	\longrightarrow W$.

	\item[Yanking]  
	\[ \Tr{X}{X}{X}(\sym_{X,X}) = \ident_{X} . \]
\end{description}
This says that $\Res$ is a \emph{traced (symmetric) monoidal category} in the 
sense of \cite{JSV} (\emph{cf.} also \cite{Has} for the symmetric and 
cartesian cases, and \cite{BE} for 
related axioms).

\subsection{Intrinsic vs. extrinsic interaction: \\ paths 
vs. names}
Why this apparent
digression into the structure of the category of resumptions? Our aim 
is to address the question of how to model \emph{interaction between 
processes}, which is surely the key notion in concurrency theory, and 
arguably in the semantics of computation as a whole. Resumptions as 
they stand model a single process in terms of its potential 
interactions with its environment. To quote Robin Milner again:
\begin{quotation}
	``A crucial feature is the ability to define the operation of \emph{binding}
	together two processes (which may represent two cooperating programs, 
	or a program and a memory, or a computer an an input/output device) 
	to yield another process representing the composite of the two 
	computing agents, with their mutual communications internalized.'' 
	\cite{Mil75}
\end{quotation}
The route Milner followed to define this binding was in terms of the 
use of ``names'' or ``labels'': in terms of resumptions, one modifies 
their defining equation to
\[ R(X, Y) = X \rightharpoonup Y \times L \times R(X, Y) \]
where $L$ is a set of labels, so that output is tagged with a label, 
which can then be used by some ``routing combinator'' to dispatch the 
output to its destination process. This led in a fairly direct line of 
descent to the action names $\alpha , \beta , \gamma$ of \textsc{ccs} 
\cite{Mil80,Mil89}, and the names of the $\pi$-calculus \cite{MPW92} 
and action structures \cite{MMP95}. Clearly a great deal has been 
achieved with this approach. Nevertheless, we wish to lodge some 
criticisms of it.
\begin{itemize}
	\item  interaction becomes extrinsic: we must add some additional 
	structure, typically a ``synchronization algebra'' on the labels 
	\cite{Win83}, which implicitly refers to some external agency for 
	matching up labels and generating communication events, rather than 
	finding the meaning of interaction in the structure we already have.

	\item  interaction becomes ad hoc: because it is an ``invented'' 
	additional structure, many possibilities arise, and it is hard to 
	identify any as canonical.

	\item  interaction becomes global: using names to match up 
	communications implies some large space in which potential 
	communications ``swim'', just as the use of references in imperative 
	languages implies some global heap. Although the scope of names may 
	be delimited, as in the $\pi$-calculus, the local character of 
	particular interactions is not immediately apparent, and must be 
	laboriously verified. This appears to account for many of the 
	complications encountered in reasoning about concurrent 
	object-oriented languages modelled in the $\pi$-calculus, as 
	reported in \cite{Jon,Jon96}.
\end{itemize}
We will now describe a construction which appears in \cite{JSV}, and 
which can be seen as a general form of the ``Geometry of 
Interaction'' \cite{Gir89}, and also as a general but basic form of 
game semantics \cite{Abr96}. This construction applies to any traced 
monoidal category $\CC$, \ie\ to any calculus of boxes and wires 
closed under
series and parallel composition and feedback, and builds a compact closed 
category $\GG 
(\CC )$, into which $\CC$ fully and faithfully embeds. (It is in fact 
the unit of a (bi)adjunction between the categories of traced 
monoidal and compact closed categories.) Its significance in the 
present context is that it gives a general way of introducing a 
symmetric notion of interaction which addresses the issues raised above:
\begin{itemize}
	\item  interaction is intrinsic: it is found from the basic idea that 
	processes are modelled in terms of their interactions with their 
	environment. Building in the distinction between ``process'' and 
	``environment'' at a fundamental level makes interaction inherent in the 
	model, rather than something that needs to be added.

	\item  interaction is modelled as composition in the category $\GG 
	(\CC )$. Thus interaction is aligned with the 
	computation-as-cut-elimination paradigm, and hence a unification of 
	concurrency 
	with other work in denotational semantics, type theory, categorical 
	logic etc. becomes possible. See \cite{AGN,Abr94,Abr95} for a 
	detailed discussion of this point.

	\item  interaction is local. The dynamics of composition traces 
	out ``information paths'', which are closely related to the types of 
	the processes which interact. There is no appeal to a global 
	mechanism for matching names. As we will see, this is general enough 
	to model $\lambda$-calculus, state and concurrency, but, we believe, 
	carries much more structure than the use of names to mediate 
	interactions.
\end{itemize}
\section{The $\GG$ construction}
Given a traced monoidal category $\CC$, we define a new category 
$\GG (\CC )$ as follows:
\begin{itemize}
	\item  The objects of $\GG (\CC )$ are pairs $(A^{+}, A^{-})$ of 
	objects of $\CC$. The idea is that $A^{+}$ is the type of ``moves by 
	Player (the System)'', while $A^{-}$ is the type of ``moves by 
	Opponent (the Environment)''.

	\item  A morphism $f : (A^{+}, A^{-}) \longrightarrow (B^{+}, B^{-})$ 
	in $\GG (\CC )$ is a morphism 
	\[ f : A^{+} \otimes B^{-} \longrightarrow A^{-} \otimes B^{+} \]
	in $\CC$.
	
	\item Composition is defined by symmetric feedback (\cf\ \cite{AJ94a,AJ94}):

	\[
	\setlength{\unitlength}{0.0125in}
\begin{picture}(361,195)(135,550)
\thicklines
\put(350,560){\line( 1, 0){ 30}}
\put(220,560){\line( 1, 0){ 40}}
\put(350,740){\vector( 1, 0){ 30}}
\put(260,740){\vector(-1, 0){ 40}}
\put(260,560){\line( 1, 2){ 90}}
\put(260,740){\line( 1,-2){ 90}}
\put(440,620){\vector( 0,-1){ 60}}
\put(440,740){\vector( 0,-1){ 60}}
\put(380,620){\vector( 0,-1){ 60}}
\put(380,740){\vector( 0,-1){ 60}}
\put(220,620){\vector( 0,-1){ 60}}
\put(160,620){\vector( 0,-1){ 60}}
\put(220,740){\vector( 0,-1){ 60}}
\put(160,740){\vector( 0,-1){ 60}}
\put(360,620){\framebox(100,60){}}
\put(140,620){\framebox(100,60){}}
\put (445,565) {\makebox(0,0) [lb] {\raisebox{0pt}[0pt][0pt]{ $C^+$}}}
\put (445,720) {\makebox(0,0) [lb] {\raisebox{0pt}[0pt][0pt]{ $C^-$}}}
\put (360,565) {\makebox(0,0) [lb] {\raisebox{0pt}[0pt][0pt]{ $\hspace{-2mm}B^-$}}}
\put (360,720) {\makebox(0,0) [lb] {\raisebox{0pt}[0pt][0pt]{ $\hspace{-2mm}B^+$}}}
\put (225,565) {\makebox(0,0) [lb] {\raisebox{0pt}[0pt][0pt]{ $\hspace{-2mm}B^+$}}}
\put (225,720) {\makebox(0,0) [lb] {\raisebox{0pt}[0pt][0pt]{ $\hspace{-2mm}B^-$}}}
\put (135,565) {\makebox(0,0) [lb] {\raisebox{0pt}[0pt][0pt]{ $A^-$}}}
\put (135,720) {\makebox(0,0) [lb] {\raisebox{0pt}[0pt][0pt]{ $A^+$}}}
\put (405,645) {\makebox(0,0) [lb] {\raisebox{0pt}[0pt][0pt]{ g}}}
\put (190,645) {\makebox(0,0) [lb] {\raisebox{0pt}[0pt][0pt]{ f}}}
\end{picture} \]
		If  $f : (A^{+}, A^{-}) \longrightarrow (B^{+}, B^{-})$ and
	$g : (B^{+}, B^{-}) \longrightarrow (C^{+}, C^{-})$ then
	$f ; g : (A^{+}, A^{-}) \longrightarrow (C^{+}, C^{-})$ is defined by
	\[ f;g = \Tr{A^{+} \otimes C^{-}}{A^{-} \otimes C^{+}}{B^{-} \otimes 
	B^{+}}(\alpha ; f \otimes g ; \gamma ) \]
	where
	\[ \alpha : A^{+} \otimes C^{-} \otimes B^{-} \otimes B^{+} \isoarrow 
	A^{+} \otimes B^{-} \otimes B^{+} \otimes C^{-} \]
	and
	\[ \gamma : A^{-} \otimes B^{+} \otimes B^{-} \otimes C^{+} \isoarrow
	A^{-} \otimes C^{+} \otimes B^{-} \otimes B^{+} \]
	are the canonical isomorphisms defined using the symmetric monoidal 
	structure. (Again, we have elided associativity isomorphisms.)
	\item The identities are given by the symmetry isomorphisms in $\CC$:
	\[ \ident_{(A^{+}, A^{-})} = \sym_{A^{+}, A^{-}} : A^{+} \otimes 
	A^{-} \isoarrow A^{-} \otimes A^{+} . \]
\end{itemize}
There is an evident involutive duality on this category, given by
\[ (A^{+}, A^{-})^{\ast} = (A^{-}, A^{+}) . \]
There is also a tensor structure, given by
\[ (A^{+}, A^{-}) \otimes (B^{+}, B^{-}) = (A^{+} \otimes B^{+}, A^{-} 
\otimes B^{-})  . \]
$\GG (\CC )$ is a compact-closed category \cite{KL}, with internal 
homs given by
\[ (A^{+}, A^{-}) \linimpl (B^{+}, B^{-}) = (A^{-} \otimes B^{+}, A^{+} 
\otimes B^{-})  . \] 
\section{Examples}
\subsection{From resumptions to strategies}
To interpret the category $\GG (\Res )$, think of an object $(X^{+}, 
X^{-})$ as a rudimentary two-person game, in which $X^{+}$ is the set 
of moves for Player, and $X^{-}$ the set of moves for Opponent. A 
resumption $f : X^{-} \longrightarrow X^{+}$ is then a 
\emph{strategy} for Player. Note that we can represent such a 
strategy by its set of \emph{plays}:
\[ P(f) = \{ x_{1}y_{1} \cdots x_{k}y_{k} \mid f(x_{1}) = (y_{1}, 
f_{1}), \ldots , f_{k-1}(x_{k}) = (y_{k}, f_{k}) \} . \]
One can then show that composition in $\GG (\Res )$ is given by 
``parallel composition plus hiding'' \cite{Abr94a,AJ94,Abr96}:
\[ P(f;g) = \{ s \restrict X, Z \mid s \in P(f) || P(g) \} \]
\[ S || T = \{ s \in \mathcal{L}(X, Y, Z) \mid s \restrict X, Y \in S 
\wedge s \restrict Y, Z \in T \} \]
where $X = X^{+} + X^{-}$, $Y = Y^{+} + Y^{-}$, $Z = Z^{+} + Z^{-}$, 
and
\[ \mathcal{L} (S_{1}, S_{2}, S_{3}) = \{ s \in (S_{1} + S_{2} + S_{3})^{\ast} 
\mid s_{i} \in S_{j} \wedge s_{i+1} \in S_{k} \; \Longrightarrow \; 
| j - k | \leq 1 \} . \]
The identities are the ``copycat'' strategies as in \cite{AJ94,Abr96}. We can 
then obtain the simple category of games described in \cite{Abr96} by 
applying a specification structure in the sense of \cite{AGN96} to 
$\GG (\Res )$, in which the properties over $(X^{+}, X^{-})$ are the 
prefix-closed subsets of $(X^{-}X^{+})^{\ast}$, \ie\ the ``safety 
properties'' \cite{AP}, which in this context are the game trees.

\subsection{Some geometries of interaction}
Suppose we begin with the simpler category $\Pfn$ of sets and partial 
functions (which is a lluf sub-category of $\Res$). This is easily 
seen to be a sub-traced-monoidal category of $\Res$, with tensor as 
disjoint union, and the trace 
given by a sum-of-paths formula (\cf\ \cite{AM}). That is, if
\[ f : X + U \rightharpoonup Y + U \]
is a partial function, then
\[ \Tr{X}{Y}{U}(f) = \bigvee_{k \in \omega} f_{k} , \]
where $f_{k}(x)$ is defined and equal to $y$ iff starting from $x$ we 
perform exactly $k$ iterations of the feedback loop around $U$ before 
exiting at Y with result $y$:
\[ f_{k} = \inl_{X,U}; (f; [0, \inr_{X,U}])^{k}; f; [\ident_{Y}, 0] \]
where $0$ is the everywhere undefined partial function. We can 
think of this sub-category of $\Res$ as the ``one-state resumptions'', 
so that, applying the $\GG$ construction to $\Pfn$ we get a category 
of history-free strategies \cite{AJ94}.

As a minor variation, we could start with the category $\PInj$ of 
sets and partial injective maps. Then $\GG (\PInj )$ is essentially 
the original Geometry of Interaction construction of Girard, as 
explained in \cite{AJ94,AJM}. In particular, the composition in $\GG 
(\PInj )$ corresponds exactly to the Execution Formula. This category 
can be lifted to the setting of Hilbert spaces by applying the free 
construction described in \cite{Barr}, which sends a set $X$ to the 
Hilbert space $l_{2}(X)$ of square summable families $\{ a_{x} \mid x 
\in X \}$.

As a final variation, we could start with $\Rel$, the category of sets 
and relations. This yields a non-deterministic version of the Geometry 
of Interaction, which can be generalized via non-deterministic 
resumptions to a category of non-deterministic strategies.
$\GG (\Rel )$  is the example mentioned at the end of \cite{JSV}.

\subsection{Stochastic interaction}
As a more substantial variation of the above, consider the following 
category of \emph{stochastic kernels} \cite{Law,Giry}. Objects are 
structures $(X, \MM (X))$, where $\MM (X)$ is a $\sigma$-algebra of 
subsets of $X$. A morphism $f : X \longrightarrow Y$ is a function
\[ f : X \times \MM (Y) \longrightarrow [0, 1 ] \]
such that for each $x \in X$ $f(x, \cdot ) : \MM (Y) \longrightarrow 
[0, 1 ]$ is a measure, and for each $M \in \MM (Y)$, $f(\cdot , 
M) : X \longrightarrow [0, 1 ]$ is a 
measurable function. One can think of stochastic kernels as 
``probabilistic transition functions''. Note that we do not require 
that each $f(x, \cdot )$ is a probability measure, \ie\ that 
$f(x,Y) = 1$, since we wish to allow for ``partial'' transition 
functions. 

Composition is by 
integration: if $f : X \rightarrow Y$ and $g : Y \rightarrow Z$, then
\[ f ; g (x, M) = \int_{Y} g(\cdot , M) df(x, \cdot ) . \]
Identities are given by point measures:
\[ \ident_{X} (x, M) = \left\{ \begin{array}{ll}
                               1, & x \in M \\
                               0, & x \not\in M .
                               \end{array} \right.
\]
Tensor product is given by disjoint union; note that $\MM (X+Y) \cong 
\MM (X) \times \MM (Y)$.

Feedback is given by a
sum-over-paths formula. Given $f : X \otimes U \longrightarrow Y 
\otimes U$, and $x \in X$, we define for each $k \in 
\omega$ a measure $\mu_{k}$ on $\MM (U)$ which gives the probability that we will 
end up in $M$ starting from $x$ after exactly $k$ traversals of the 
feedback loop:
\[ \mu_{0}(M)  =  f(\inl (x), (\varnothing , M)) \]
\[ \mu_{k+1}(M)  =  \int_{U} f(\inr (\cdot ), (\varnothing , M)) d 
\mu_{k} . \]
The probability that we will end up in $M \in \MM (Y)$ starting from 
$x$ after exactly $k$ iterations of the feedback loop is given by:
\[ f_{0}(x,M)  =  f(\inl (x), (M, \varnothing )) \]
\[ f_{k+1}(x,M)  =  \int_{U} f(\inr (\cdot ), (M, \varnothing )) 
d\mu_{k} . \]
Finally, the trace is defined by summing over all paths:
\[ \Tr{X}{Y}{U}(f)(x,M) = \Sigma_{k \in \omega} f_{k}(x,M). \]
\subsection{From particles to waves: the ``New Foundations'' version of 
Geometry of Interaction}
All the above models can be thought of as dynamical systems in which 
an information ``token'' or ``particle'' traces some path around a 
network. This particulate interpretation of diagrams of boxes and 
wires is supported by the ``additive'' (disjoint union) 
interpretation of the tensor. It is also possible to give an 
interpretation in which an information  ``wave'' travels through the 
network; formally, this will be supported by a ``multiplicative'' 
(cartesian product) interpretation of the tensor.

Specifically, we can define a traced monoidal structure on the 
category $\Cpo$ of cpo's and continuous functions, in which the tensor 
is given by the  cartesian product, and feedback by the least fixpoint 
operator: that is, if $f : D \times A \longrightarrow E \times A$, then
\[ \Tr{D}{E}{A}(f) = \lambda d : D. \, f (d, \YY (f(d, \cdot ) ; \snd)) ; 
\fst . \]
The category $\GG (\Cpo )$ is then exactly the category $\GI (\CC )$ 
described in \cite{AJ94a}.

A sub-category of this category will consist of dataflow networks, 
built up from objects which are domains of streams. The symmetric 
feedback operator giving the composition in $\GG (\Cpo )$ has been 
used in this context  \cite{SDW96,Sto},  \emph{inter alia} in developing 
assumption/commitment style proof rules for dataflow networks.
\subsection{The continuous case?}
One final ``example'' should be mentioned, although we have not as yet 
succeeded in working out the details. The operations of series and 
parallel composition and feedback are standard in continuous-time 
control systems, electronic circuits and analogue computation. In 
particular, feedback is interpreted by solving a differential 
equation. There should then presumably be a traced monoidal 
category $\CC$
of manifolds and smooth maps, for which $\GG (\CC )$ would give an
``infinitesimal'' model of interaction. Such a category might be 
relevant to the study of hybrid systems \cite{Hyb}.
\section{Consequences}
We shall, very briefly, sketch some further developments from this point.
\subsection{Correctness issues}
We can associate correctness properties with the rudimentary types 
of $\GG (\CC )$, in the setting of specification structures 
\cite{AGN96}. Types can then carry strong correctness information, 
and the type inference rule for composition
\[ \frac{f : A \rightarrow B \quad g : B \rightarrow C}{f;g : A 
\rightarrow C} \]
becomes a compositional proof rule for process interaction. See 
\cite{Abr94,Abr95,AGN,AGN96} for further discussion and applications.

We shall mention some particular cases for the examples described 
above.

\paragraph{Resumptions}
In this case, we can get the structure of games as safety properties, 
and of winning strategies as liveness properties, as described in 
\cite{AJ94,Abr96}. In particular, the fact that winning strategies 
are closed under composition corresponds to a guarantee that there is 
no ``infinite chattering'' \cite{Hoa85} in interaction.

\paragraph{Geometry of Interaction}
In this case, we can focus on nilpotency as  a semantic analogue of 
normalization, as in \cite{Gir89}, or instead proceed as in the 
previous example, as in \cite{AJ94}, where a Full Completeness 
Theorem for Multiplicative Linear Logic is obtained.

\subsection{Modelling types and functions}
The divide between concurrency theory and denotational 
semantics, type theory and categorical logic is bridged in our 
approach, since the categories we construct, or derivatives thereof, 
have the right structure to model typed, higher-order programming 
languages. The key point is that we are now modelling functions as 
processes, and function application as a particular form of process 
interaction, as advocated in \cite{Mil92}, but in a highly structured, 
syntax-free and compositional fashion.

Moreover, the quality of these process models of functional 
computation is high: the models based on games yielded the first 
syntax-independent constructions of fully abstract models for PCF 
\cite{AJM,HO}, and this has been followed by a number of further 
results \cite{AM95,M96a,M96b}. The degree of mathematical 
structure in these models is also witnessed by the axiomatic treatment 
of full abstraction
it has been possible to extract from them \cite{Abr96b}.
\subsection{State and concurrency}
It has also proved possible to give a game semantics for Idealized 
Algol \cite{Abr95a}, which is a clean integration of higher-order 
functional programming with imperative features and block structure 
\cite{Rey,Ten}. 
Again, this has led to the first syntax-independent construction of a 
fully abstract model \cite{AM96}. The treatment of local variables 
is process-based, following the line of \cite{Mil75,Mil80, Red96}; 
but with the the right mathematical tools now available, a more 
definitive treatment can be given, as confirmed by the results on full 
abstraction.

This model of Idealized Algol extends smoothly to incorporate 
concurrency \cite{Abr95a}. It remains to be seen how accurate the
model of the concurrent language is, but the situation looks quite 
promising: moreover, Idealized Parallel Algol is rich enough to represent rather 
directly many of the features of today's concurrent object-oriented 
languages.

\bibliography{biblio,biblio1}
\end{document}